\documentclass[aps,twocolumn,prb,showpacs]{revtex4}

\usepackage[dvips]{color}
\usepackage{graphicx}
\usepackage{epsfig}
\usepackage{dcolumn}
\usepackage{color}
\usepackage{amssymb}
\usepackage[english]{babel}
\usepackage{bm}
\usepackage{amsmath}
\newcommand{\eq}[1]{(\ref{eq:#1})}
\newcommand{\eqname}[1]{\label{eq:#1}}

\begin{document}

\title{Analog Hawking radiation from an acoustic black hole in a flowing polariton superfluid}

\author{Dario Gerace }
\affiliation{Dipartimento di Fisica, Universit\`a di Pavia,  via Bassi 6, I-27100 Pavia, Italy}
\email{dario.gerace@unipv.it}

\author{Iacopo Carusotto }
\affiliation{ INO-CNR BEC Center and Dipartimento di Fisica, Universit\`a di Trento, I-38123 Povo, Italy  }
 \email{carusott@science.unitn.it}

\pacs{
03.70.+k; 
42.50.Lc; 
42.50.Ct 
}

\date{\today}

\begin{abstract}
We theoretically study Hawking radiation processes from an analog acoustic black hole in a flowing superfluid of exciton-polaritons in a one-dimensional
semiconductor microcavity. Polaritons are coherently injected into the microcavity by a laser pump with a suitably tailored spot profile. An event horizon with a large analog surface gravity is created by inserting a defect in the polariton flow along the cavity plane.
Experimentally observable signatures of the analog Hawking radiation are identified in the scattering of phonon wavepackets off the horizon, as well as in the 
spatial correlation pattern of quantum fluctuations of the polariton density. The potential of these table-top optical systems as analog models of gravitational 
physics is quantitatively confirmed by numerical calculations using realistic parameters for state-of-the-art devices.
\end{abstract}

\maketitle

\section{Introduction}
The quantum mechanical properties of the vacuum state in quantum field theories are presently attracting the attention of researchers from very different communities, from astrophysics and gravitation, to quantum optics and condensed matter physics. A central idea of all these studies is the possibility of converting zero-point quantum fluctuations into observable quantum vacuum radiation by some spatial and/or temporal dependence of the background over which the quantum field is propagating~\cite{birrell,milonni}. A number of fundamental physical effects belong to this category, from the dynamical Casimir effect when the boundary conditions of the quantum field are rapidly varied in time~\cite{DCE,DCE_rev}, to the Hawking radiation when it propagates on a curved space-time showing a black hole horizon~\cite{hawking74,hawking75}.

In the wake of the pioneering work by Unruh~\cite{unruh81}, researchers have started investigating condensed-matter systems where the propagation of some low-energy excitation field follows an effectively curved space-time geometry. In suitable configurations showing a black hole horizon for this low-energy excitation, a quantum vacuum emission is expected to appear via a mechanism analogous to Hawking radiation. Among the many systems that have been investigated in this perspective~\cite{novello,LivRev}, low-temperature superfluids and nonlinear optical systems are nowadays considered the most promising ones.

Dilute superfluids such as Bose-Einstein condensates of ultracold atoms join a very simple spectrum of elementary excitations with sonic dispersion, with the possibility of pushing the sensitivity of measurements down to the quantum limit. Their potential as analog models was first proposed in Refs.~\onlinecite{zoller,fedichev,liberati} and later confirmed by {\em ab initio} numerical simulations of the condensate dynamics~\cite{carusott2008njp}. Even if acoustic black hole configurations have been experimentally realized~\cite{steinhauer}, no evidence of Hawking radiation has been reported yet.

Analog models based on nonlinear optical systems were pioneered in Ref.~\onlinecite{philbin2008}: a strong pulse of light propagating in a Kerr nonlinear medium can be used to generate a moving spatial interface separating regions of different light velocity. For suitably chosen parameters, the interface then behaves as a horizon that can be crossed by light in one direction only, which should result in Hawking radiation being emitted. Some experimental claim in this direction was recently reported~\cite{faccio2010} and has raised a number of interesting questions concerning the interpretation of the observed radiation in terms of Hawking processes~\cite{faccio2010comment,liberati2011,finazzi2012}.

A completely new perspective to analog models based on nonlinear optical systems was opened by Marino~\cite{marino2008}, who first proposed the use of quantum fluids of light to generate acoustic black hole configurations and, then, to look for the Hawking radiation of Bogoliubov phonons on top of the photon fluid. 
In the following years, this idea has been pushed forward by a number of authors, who have considered different geometrical configurations~\cite{marino2009,fouxon2010} and material systems, and very recently exciton-polariton fluids in semiconductor microcavities~\cite{malpuech2011prb}.

In the present work, we report a comprehensive study of analog Hawking radiation effects in quantum fluid of light, specifically exploiting the quantum fluid properties of an exciton-polaritons condensate. 
Our theoretical model fully includes their intrinsically non-equilibrium nature~\cite{ICCC_RMP} and describes quantum fluctuations of the polariton field within the so-called truncated-Wigner formalism of degenerate quantum gases~\cite{sinatra,carusott2005prb}. 
As a result, our numerical calculations are able to provide quantitative predictions for the observable quantities, and to point out clear and accessible signatures of analog Hawking radiation in the emitted light from the cavity. Even if our discussion is mostly focused on the specific case of semiconductor microcavity devices~\cite{Weisbuch,varenna_proc,ciutiSST,ICCC_RMP} where polariton superfluidity has been first demonstrated~\cite{amo2009nat}, all our conclusions straightforwardly extend to generic planar cavity devices filled with or made of a Kerr nonlinear medium. 

The article is organized as follows. In Sec.~\ref{theory} we introduce the physical system under consideration and we summarize the theoretical tools that are used to describe it. In Sec.~\ref{horizon} we describe the original laser beam configuration that we propose to generate an analog black hole horizon with a large surface gravity. The observable consequences of Hawking mode-conversion processes are illustrated in the following sections. In Sec.~\ref{scattering} we discuss the scattering of a coherent phonon wavepacket on the horizon: the signature of classical Hawking processes is visible as an additional wavepacket emerging from the horizon on a negative-norm branch. In Sec.~\ref{Hawking}, we present the numerical evidence of the Hawking radiation due to the conversion of zero-point quantum fluctuations into observable radiation, and we compare it with the theoretical expectations. Conclusions are finally drawn in Sec.~\ref{concl}.

\section{Physical system and theoretical model}\label{theory}

\begin{figure}[t]
\begin{center}
\vspace{-0.5cm}
\includegraphics[width=0.45\textwidth]{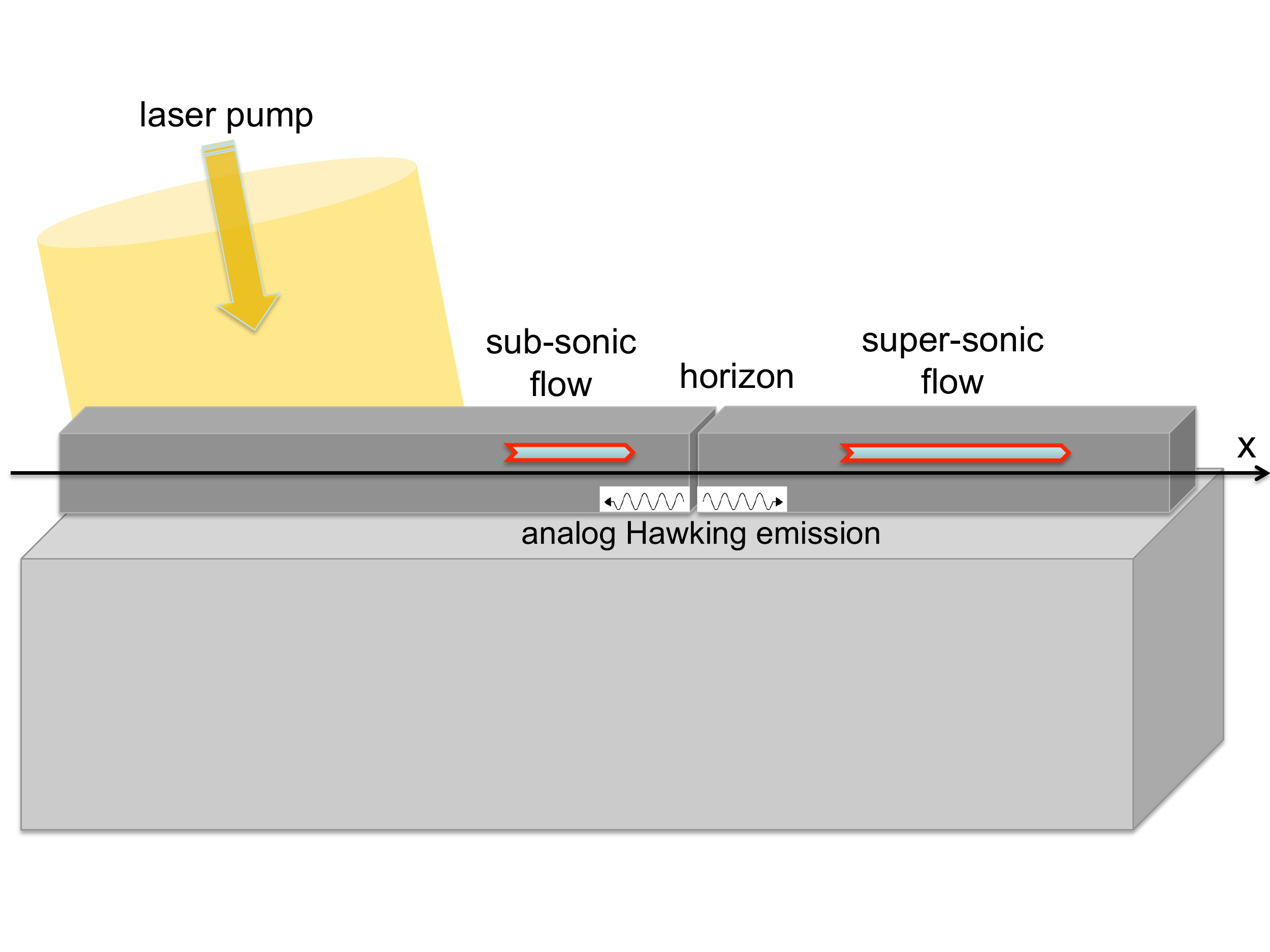}\\
\vspace{-0.8cm} 
\includegraphics[width=0.48\textwidth]{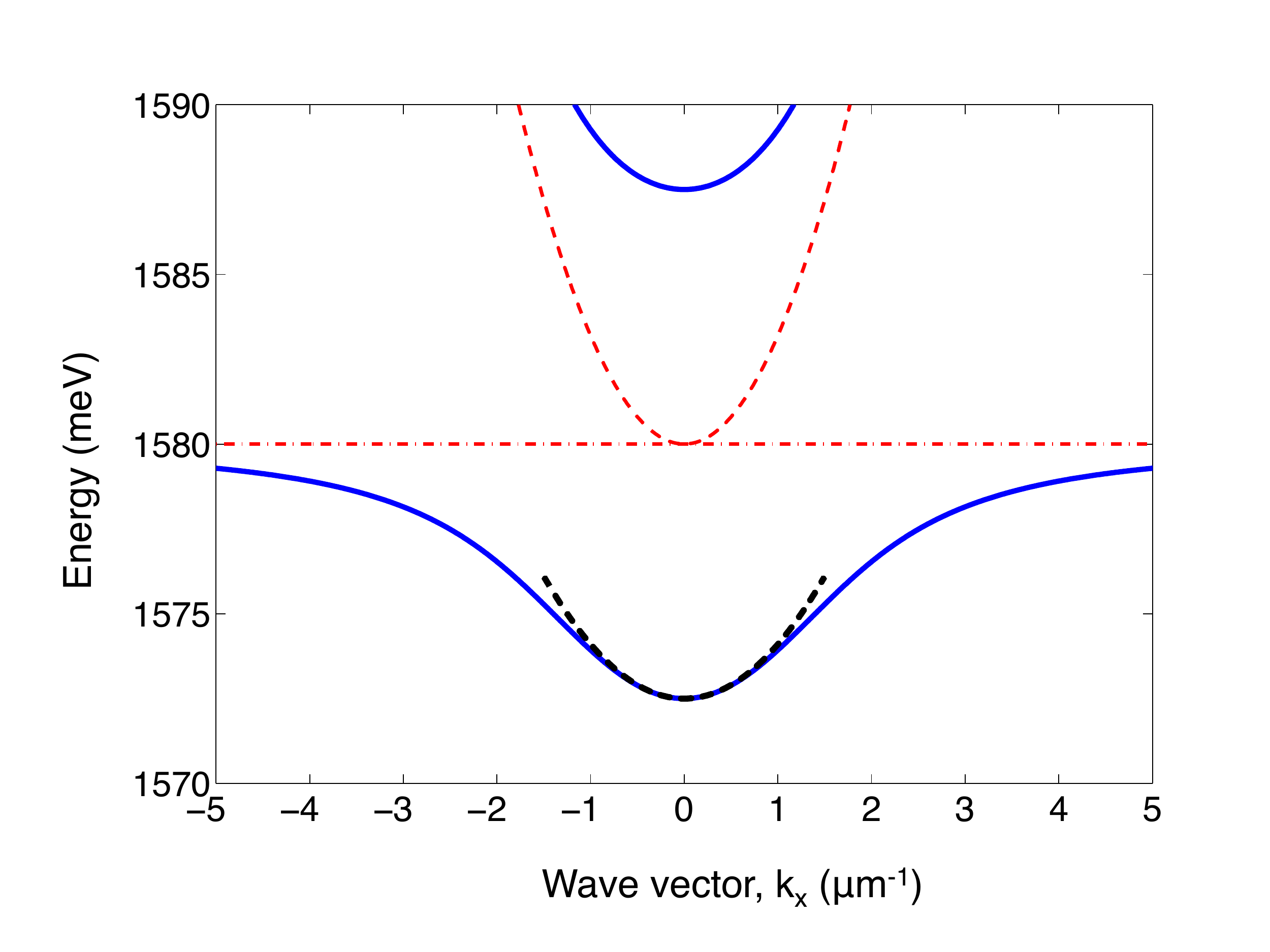}\\
\caption{(Color online) Upper panel: sketch of the polariton wire device under consideration.
Lower panel: energy-dispersion of the upper and lower polariton branches as a function of the wavevector $k_x$ along the wire 
axis. The thin dashed and dash-dotted lines indicate the bare cavity photon and exciton dispersions, respectively. The thick dashed line indicates the parabolic approximation of the lower polariton dispersion. 
System parameters are inspired from Ref.~\onlinecite{wertz2010}:  $\hbar\omega_{\mathrm{x}}^{(o)}=\hbar\omega_{\mathrm{c}}^{(o)}=1580$~meV,
   $\hbar\Omega_R=7.5$~meV, $m_{\mathrm{c}} =1.2\times 10^{-5} m_0$ (with $m_0$ the free electron mass), $m_{\rm x}\gg m_{\rm c}$. } \label{fig1}
\end{center}
\end{figure}

We consider a polariton wire device where cavity photons propagating along one dimension are resonantly coupled to the fundamental exciton transition in one or more quantum wells embedded in the cavity layer (in-plane polarized heavy-hole to conduction band transition). 
Starting from a planar microcavity where light is confined along the growth axis $z$ by a pair of distributed Bragg reflectors, a suitable etching procedure is used to laterally pattern the device in the $y$ direction, and create the ridge structure that is schematically shown in Fig.~\ref{fig1}(a). Provided the losses are weak enough, the bosonic excitations that result from the strong coupling of the cavity photon with the quantum well exciton have the typical mixed light-matter nature of one-dimensional exciton-polaritons, free to propagate along the wire axis according to the dispersion law that is plotted in Fig.~\ref{fig1}(b). A high quality experimental realization of this polariton wire concept was reported in Ref.~\onlinecite{wertz2010}, from which we shall take the state-of-the art system parameters to be used in the numerical simulations.

\subsection{The quantum field Hamiltonian}

A theoretical description of the dynamics of this system can be developed in terms of the standard Hamiltonian for the two coupled bosonic fields describing the quantum well exciton and the cavity photon~\cite{ciutiSST}
\begin{multline}
\eqname{model}
\mathcal{H} =  \int \mathrm{d}x \, 
\vec{\Psi}^{\dagger}(x) H_0 \vec{\Psi} (x)+  \\
+  \frac{\hbar g}{2} \int \mathrm{d}x \, \hat{\psi}_{\mathrm{x}}^{\dagger} (x)  \hat{\psi}_{\mathrm{x}}^{\dagger} (x) 
\hat{\psi}_{\mathrm{x}} (x) \hat{\psi}_{\mathrm{x}} (x)  \\
+  \int \mathrm{d}x \, \hbar  E(x,t)   \hat{\psi}_{\mathrm{c}}^{\dagger} (x)  + \mathrm{h.c.}\, ,
\end{multline}
where ${x}$ is the longitudinal coordinate along the wire axis and the two-component operator vector $\vec{\Psi} = (\hat{\psi}_{\mathrm{x}},\hat{\psi}_{\mathrm{c}})^T$ summarizes the quantum fields describing the exciton $\hat{\psi}_{\rm x}(x)$ and photon $\hat{\psi}_{\rm c}(x)$ fields. Each of them satisfies one-dimensional bosonic commutation rules, e.g. 
$[\hat{\psi}_{i}(x),\hat{\psi}^{\dagger}_{j}(x')]=\delta_{i,j}\delta(x-x')$, with $i,j=\{{\rm x},{\rm c}\}$. Throughout the whole paper, we restrict ourselves to the case where the system is pumped on a single spin state, so that the spin degrees of freedom can be neglected in the theoretical model.

The single-particle Hamiltonian $H_0$ describing the evolution of the non-interacting exciton and cavity photon fields has the simple representation
\begin{equation}
\label{eq:single_ham}
H_0 = 
\left(  \begin{array} {cc}
\hbar \omega_{\mathrm{x}}({-i  \partial_x}) +  V_{\mathrm{x}}(x) & \hbar \Omega_R \\
\hbar \Omega_R  &  \hbar  \omega_{\mathrm{c}}({-i \partial_x}) + V_{\mathrm{c}}(x) 
\end{array}\right) \, ,
\end{equation}
in terms of the bare cavity photon and exciton dispersion law $\omega_{\rm c,x}(k_x)\approx \omega_{\mathrm{c,x}}^{(o)}+ \hbar^2  {k_x}^2 /(2 m_{\mathrm{c,x}}) $ in a spatially homogeneous one-dimensional system. The rest energy $\omega_{\rm c}^{(o)}$ and the effective mass $m_{\mathrm{c}}$ of the cavity photon are determined by the spatial confinement along the $z$ and $y$ axis by the DBR mirrors and by the total internal reflection at the etching interfaces, respectively. In practical calculations the exciton dispersion can be safely neglected, $\omega_{\rm x}(k_x)=\omega_{\rm x}^{(o)}$, as the exciton mass $m_{\rm x}$ is orders of magnitude larger than the photon mass. The strength of the light-matter coupling is quantified by the Rabi frequency $\Omega_R$, proportional to the square root of the oscillator strength per unit area of the excitonic transition~\cite{varenna_proc}.

In spatially homogeneous systems, for which the external potentials acting on the cavity photon and the quantum well exciton vanish ($V_{\mathrm{x},\mathrm{c}}(x)=0$), diagonalization of the single particle Hamiltonian $H_0$ as a function of wave vector $k_x$ leads to the well-known upper and lower polariton branches of dispersions $\omega_{UP}(k_x)$ and $\omega_{LP}(k_x)$. 
As it is shown in Fig.~\ref{fig1}(b) the bottom part of the lower polariton dispersion is accurately approximated by a parabolic form
\begin{equation}
\omega_{LP}(k_x)\simeq \omega_{LP}^{(o)}+\frac{\hbar k_x^2}{2m_{LP}}
\end{equation}
where the  $\omega_{LP}^{(o)}$ and $m_{LP}$ parameters have the clear physical meaning of a rest energy and an effective mass for the (lower) polariton. In the resonant case $\omega_{\rm x}^{(o)}=\omega_{\rm c}^{(o)}$, one has the simple relation $m_{LP}\simeq 2m_c$. 

Several methods to generate external potentials $V_{\rm x,c}(x)$ with almost arbitrary spatial shapes along the $x$ direction have been experimentally demonstrated in the last years, from the in-plane patterning of the microcavity layer~\cite{pattern}, to the all-optical potential created by a strong pump laser with opposite polarization~\cite{Amo}, to the application of a mechanical stress to the device~\cite{Balili}.

Exciton-exciton interaction, due to the Coulomb interactions between the electron and holes forming the excitons, are the main source of optical nonlinearity in this problem. Experimental measurements~\cite{wertz2011,tanese2012} report a value on the order of $\hbar g_{2D}\approx 2.5 \, \mu\textrm{eV}\cdot\mu\textrm{m}^2$, which translates into a reduced one-dimensional interaction parameter to be used in the Hamiltonian \eq{model} $g=g_{2D}\int_{{\rm w}} \mathrm{d}y |\Phi_x(y)|^4$, where the integral is over the transverse wire width ($-L_{\rm w}/2 < y < L_{\rm w}/2$), and the forth power of the exciton wave function comes from the Kerr-type nature of the nonlinearity. Assuming the transverse confinement of the exciton envelope function to be described by $\Phi_x(y)=\sqrt{2/L_{\rm w}} \cos{(\pi y/L_{\rm w})}$, a good estimate can be given to be $g=6 g_{2D} /L_{\rm w}$. Taking into account the exciton weight in the polariton, this translates in the resonant $\omega_{\rm x}^{(0)}=\omega_{\rm c}^{(0)}$ case into a one-dimensional polariton-polariton interaction constant $g_{LP}\simeq g/4$.

Polaritons are coherently injected into the system by an external laser drive coupled to the cavity photon via the non-perfect reflectivity of the DBR mirrors. In the theoretical model, such processes are described by the terms in the last row of Eq.~\eq{model}, where $E(x,t)$ is the spatio-temporal profile of the coherent pump beam including the coupling coefficient proportional to the DBR transmission amplitude. 
Correspondingly, photons (excitons) are subjected to radiative losses at a rate $\gamma_{\rm c}$ ($\gamma_{\rm x}$) that have to be described at the level of the master equation for the density matrix, $\rho$, for the coupled quantum fields
\begin{equation}
\frac{d\rho}{dt}=\frac{1}{i\hbar}[\mathcal{H},\rho]+\mathcal{L}[\rho].
\end{equation}
Performing the standard Born-Markov approximation~\cite{QO}, the dissipation superoperator can be cast into the Lindblad form
\begin{multline}
\mathcal{L}=\sum_{i=\mathrm{x},\mathrm{c}}
\frac{\gamma_{i}}{2}\,\int\!dx\left[
\hat{\psi}_{{i}}(x)\,\rho \hat{\psi}_{{i}}^{\dagger} (x)-
 \hat{\psi}_{{i}}^{\dagger} (x)\,\hat{\psi}_{{i}}(x)\,\rho \right.\\ -\left.
\rho \,\hat{\psi}_{{i}}^{\dagger} (x)\,\hat{\psi}_{{i}}(x)\right]  \, .
\end{multline}
The resulting decay rate of polaritons is a weighted average of the exciton and photon ones $\gamma_{\rm x,c}$: in the vicinity of the bottom of the lower polariton branch of Fig.~\ref{fig1}(b), it has an almost momentum-independent value $\gamma_{LP}\approx (\gamma_{\rm x}+\gamma_{\rm c})/2$.

\subsection{The generalized Gross-Pitaevskii equation}\label{sec:meanfield}

In a regime of weak exciton-exciton interactions, the dynamics of the system can be accurately captured by the mean-field approximation: the quantum fields $\hat{\psi}_{\rm x,c}$ are replaced by the classical fields corresponding to their expectation values
\begin{equation}
\phi_{\rm x,c}(x)=\langle \hat{\psi}_{\rm x,c}(x) \rangle
\end{equation}
and evolving according to the pair of nonlinear partial differential equations
\begin{multline}\label{eq:mf_gpe}
i\hbar \frac{d}{dt} \left( 
\begin{array}{c} \phi_{\mathrm{x}} (x,t) \\ 
 \phi_{\mathrm{c}} (x,t) \end{array} 
 \right) =  \left( 
\begin{array}{c} 0 \\ 
 \hbar E(x,t) \end{array} 
 \right) + \\
 + \left[ H_0 + \left( 
 \begin{array}{cc} \hbar g |\phi_{\mathrm{x}} (x,t) |^2 -i\hbar \frac{\gamma_{\mathrm{x}}}{2} & 0 \\ 
   0  & -i\hbar \frac{\gamma_{\mathrm{c}}}{2} \end{array} 
 \right) \right] \left( 
\begin{array}{c} \phi_{\mathrm{x}} (x,t) \\ 
 \phi_{\mathrm{c}} (x,t) \end{array} 
 \right)  \, ,
\end{multline}
which generalize to the non-equilibrium context of polaritons the well-known Gross-Pitaevskii equation (GPE) of dilute Bose condensed gases~\cite{pitaevskii_stringari}.

\label{sec:homog}
In the simplest case of a spatially homogeneous system under a coherent pump in a plane wave form 
\begin{equation}
E(x,t)=F_p\,\exp[i(k_p x - \omega_p t)]
\end{equation} 
of frequency $\omega_p$, wavevector $k_p$ and amplitude $F_p$, the analog of the equation of state can be obtained by simply injecting the same plane wave ansatz 
$\phi_{\mathrm{x},\mathrm{c}}(x,t)=  \phi^{ss}_{\mathrm{x},\mathrm{c}}\, \exp[i(k_{\mathrm{p}} x - \omega_{\mathrm{p}})t]$
into the generalized GPE \eq{mf_gpe}. 

\begin{figure}[t]
\begin{center}
\vspace{-1cm}
\includegraphics[width=0.5\textwidth]{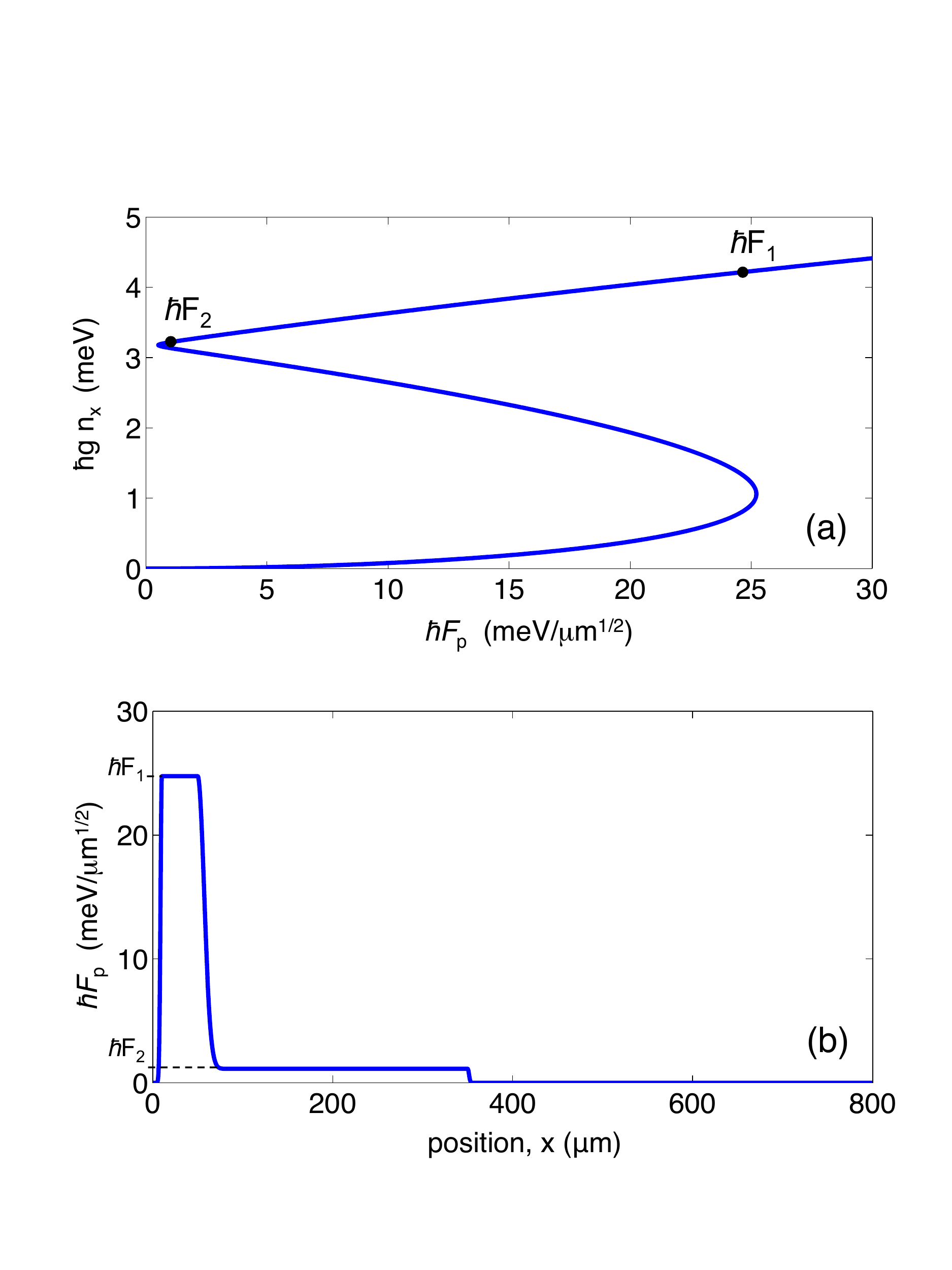}
 \vspace{-1cm}
\caption{(Color online) 
Upper panel: hysteresis loop in the polariton density vs. incident laser amplitude for a plane wave pump of wavevector $k_{\mathrm{p}}=0.2$ $\mu$m$^{-1}$, and frequency $\hbar \omega_{\mathrm{p}}=1574$ meV tuned slightly above the lower polariton branch at $\omega_{LP}(k_{\rm p})\simeq 1572.56$~meV. Equal loss rate are assumed
for the exciton and the photon fields, $\hbar\gamma_{{\rm x},{\rm c}}=0.02$ meV. 
The value $\hbar g=5\,\mu\textrm{eV}\cdot \mu\textrm{m}$ of the nonlinearity corresponds to a $L_{\rm w}=3\,\mu$m wide polariton wire (see text).
Lower panel: spatial intensity profile of the coherent pump along the wire axis,
with excitation parameters: $\hbar F_{1}=24.7487\,\textrm{meV}\cdot \mu\textrm{m}^{-1/2}$, $\hbar F_{2}=1.1314\,\textrm{meV}\cdot \mu\textrm{m}^{-1/2}$,
$k_{\mathrm{p}}=0.2\,\mu\textrm{m}^{-1}$,  $\hbar \omega_{\mathrm{p}}=1574\,\textrm{meV}$. 
} \label{fig2}
\end{center}
\end{figure}

This leads to the simple form 
\begin{multline} \label{solution_stability}
\left| \left( \Delta_{\mathrm{c}\mathrm{p}}  - i \frac{\gamma_{\mathrm{c}}}{2} \right) \left( \Delta_{\mathrm{x}\mathrm{p}}  
+ g  n_{\rm x}  - i \frac{\gamma_{\mathrm{x}}}{2} \right) -\Omega_R^2 \right|^2
n_{\rm x} =  \\ =\Omega^2_R \, |F_{\mathrm{p}}|^2  \, ,
\end{multline}
for the steady-state excitonic density $n_{\rm x}=|\phi_{\rm x}^{ss}|^2$ as a function of the pump parameters, and to an analogous expression for the cavity-photon density~\cite{carusotto2004prl,ciuti2005pss}. The pump detuning from the cavity photon and exciton modes at wavevector $k_{\rm p}$ are indicated here as $\Delta_{\rm cp}=\omega_{\rm c}(k_p)-\omega_p$ and  $\Delta_{\rm xp}=\omega_{\rm x}(k_p)-\omega_p$. 

In the quantum fluid language, this equation plays the role of the equation of state for the polariton fluid under a coherent pump in a plane wave form: while the polariton flow speed is controlled by the incident wavevector $k_{\rm p}$, the density has a more complex dependence on the pump intensity $|F_p|^2$ and frequency $\omega_p$. A most remarkable example is illustrated in Fig.~\ref{fig2}a: the choice of a pump frequency slightly above the lower polariton branch leads to an optical bistable behavior.
 
For reasons that will soon become clear, we shall concentrate our attention onto the case where the pump intensity is tuned in the close vicinity of the end-point of the upper branch of the hysteresis loop (indicated as $F_2$ in Fig.~\ref{fig2}a). In order to reach this working point in a given spatial region, a pump with a non-trivial spatial envelope, $F_p(x)$, multiplying the plane wave can be used, as shown in the lower panel of Fig.~\ref{fig2}(b). Upstream of the region of interest (i.e. for $10\,\mu\textrm{m}<x<50\,\mu\textrm{m}$ the pump amplitude is tuned to a value $F_1$ higher than the switch-on point of the hysteresis loop (i.e. the end point of the lower branch), so to switch the system to the upper branch of the hysteresis loop~\cite{shelykh2008prl}. In the long region of interest ($50\,\mu\textrm{m} < x < 350\,\mu\textrm{m}$), the pump amplitude is maintained at a spatially constant value $F_2$, so to keep the system sits at the desired working point with the desired in-plane flow wavevector $k_{\rm p}$.  

The behavior of the density $n(x)$ and the wavevector $k(x)$ in the downstream region ($350\,\mu\textrm{m}<x$) is determined by a complex interplay of propagation and losses. First numerical and analytical studies of this regime have appeared in Refs.~\onlinecite{pigeon2011prb} and \onlinecite{kamchatnov2011}: the spatial drop of the polariton density away from the pump results in an increase of the polariton speed under the effect of the interactions. The lower the losses, the slower this acceleration effect.

\subsection{Bogoliubov dispersion of collective excitations}
\label{sec:bogoliubov}

The dispersion of collective excitations on top of the steady state of the coherently pumped condensate is obtained as usual by linearizing the GPE around the steady state solution~\cite{pitaevskii_stringari}. A complete discussion of the many different behaviors that are possible depending on the pump frequency and intensity can be found in Refs.~\onlinecite{carusotto2004prl,ciuti2005pss}. Here, we shall restrict our attention to the two cases of present interest, namely (i) a coherent pump whose intensity is tuned right at the turning point of the upper branch of the hysteresis loop, and (ii) the ballistic flow regime in the downstream region.

If the polariton population is restricted to the lower polariton branch and we perform a parabolic approximation of the branch bottom with effective mass $m_{LP}$, the dispersion of the collective excitations on top of a condensate of polariton density $n_{LP}$ in uniform flow at speed $v_{LP}$ has, in both cases, the usual Bogoliubov form 
\begin{multline}
\omega_{\rm Bog}(k_x) - \omega_p \simeq \\
\sqrt{\frac{\hbar (k_x - k_{LP})^2}{2 m_{LP}} \left(\frac{\hbar (k_x - k_{LP})^2}{2 m_{LP}} + 
g_{LP} n_{LP} \right)} \\ + v_{LP} (k_x - k_{LP})+ \frac{i\,\gamma_{LP}}{2} \, .
\eqname{sonic}
\end{multline}
In the first (i) case, the local wavevector $k_{LP}$ is fixed to the in-plane component $k_p$ of the pump wavevector. In the second (ii) case, hydrodynamic effects~\cite{malpuech2011prb,pigeon2011prb,kamchatnov2011} make the local wavevector $k_{LP}(x)$ to slowly vary in space.
In both cases, the local in-plane flow speed $v_{LP}$ is related to the local wavevector by $v_{LP}=\hbar k_{LP} /m_{LP}$~\footnote{Restricting to the bottom of the lower polariton branch, it is easy to see that the polariton density can be expressed as $n_{LP}=|\phi_{\rm x}|^2+|\phi_{\rm c}|^2$, and the polariton speed is defined as $v_{LP}=j_{LP}/n_{LP}$ in terms of the polariton current $j_{LP}=\frac{1}{4i m_{\rm c}}[\hat{\psi}^*_{LP}\,\nabla {\psi}_{LP}- \psi_{LP}\,\nabla{\psi}^*_{LP}] \simeq  j_{\rm c}=\frac{1}{2i m_{\rm c}}[\hat{\psi}^*_{\rm c}\,\nabla {\psi}_{\rm c}- \psi_{\rm c}\,\nabla{\psi}^*_{\rm c}]$. Given its very heavy mass, $m_{\rm x}\gg m_{\rm c}$, the exciton fraction does not significantly contribute to the polariton current.}.

For small wave vectors, $|k_x - k_p|\, \xi \ll 1$, the Bogoliubov dispersion tends to a (lossy) sonic dispersion 
\begin{equation}
\omega_{\rm Bog}(k_x) - \omega_p \simeq c_{s}\,|k_x - k_p| + v_{LP} (k_x - k_p) + i\frac{\gamma_{LP}}{2}
\end{equation}
with a sound speed $c_{s}$ of fluid excitations such that $m_{LP}c^2_{s}=\hbar g_{LP} n_{LP}$, and a Doppler shift term due to the background fluid flow at $v_{LP}$. For large wavevectors, it recovers a (lossy) parabolic single-particle dispersion
\begin{multline}
\omega_{\rm Bog}(k_x) - \omega_p \simeq \frac{\hbar (k_x - k_p)^2}{2m_{LP}} + v_{LP} (k_x-k_p) + \\ + g_{LP} n_{LP} + i\frac{\gamma_{LP}}{2}
\end{multline}
with a Hartree energy shift $\hbar n_{LP} g_{LP}$ due to the interactions with the condensate. Plots of the Bogoliubov dispersion at different spatial positions are shown in the small panels of Figs.~\ref{fig:noBH_scheme} and \ref{fig:BH_scheme}. In the following of the paper, the will be used to interpret the numerical results for the evolution of excitation wavepackets and for the density-density correlations.

Before proceeding, it is important to emphasize the conditions underlying the dispersion \eq{sonic} of the collective excitations. In the coherent pump case (i), this form only holds at the end point of the upper branch of the hysteresis loop, while gaps and/or band sticking regions appear in all other cases because of the coherent pump locking the condensate phase~\cite{carusotto2004prl,ciuti2005pss}.

On the other hand, the presence of a soft excitation branch of sonic nature is guaranteed in the ballistic case (ii) by the fact that the condensate phase is completely free to evolve both in space and time~\cite{pigeon2011prb}. However, as the density is spatially varying, the very concept of dispersion relation holds only locally and is limited to excitations whose wavelength is small as compared to the characteristic length scale of the density profile.

\subsection{Truncated Wigner method}\label{sec:montecarlo}

A standard technique to go beyond the mean-field approximation and include the fluctuations of the polariton field around its mean value is based on the Wigner representation of quantum fields. Within the {\em truncated Wigner approximation}, the dynamics of the quantum field problem can be described by a stochastic partial differential equation that can be numerically solved.
This technique was first introduced in the context of quantum fluids in Ref.~\onlinecite{steel1998} and its application to atomic condensates was fully developed and characterized in Ref.~\onlinecite{sinatra}. Its extension to the polariton case was developed in Ref.~\onlinecite{carusott2005prb}: remarkably, the presence of loss and pump terms in the stochastic equations suppresses some of the difficulties of the truncated Wigner method and guarantees its accuracy as long as interactions between individual polaritons are weak enough. If a spatial grid of real-space spacing $\Delta x$ is used to numerically solve the stochastic partial differential equation, the weak interaction condition can be formulated as
$|g| \ll \gamma\, \Delta x$.
While this condition rules out the possibility of using the truncated Wigner method to study strongly correlated polariton states, e.g. in the polariton blockade regime~\cite{verger2006,gerace2009,carusott2009prl,ferretti2010}, it does not hinder the study of quantum hydrodynamics effects such as the analog Hawking radiation, which originate from the collective dynamics of a large number of polaritons. 

The stochastic partial differential equations of the truncated Wigner method for the coupled exciton and polariton fields has the explicit form~\cite{carusott2005prb}
\begin{widetext}
\begin{equation}
\label{eq:wigner}
i\hbar \left( 
\begin{array}{c} d\phi_{\mathrm{x}}\\ 
 d\phi_{\mathrm{c}} \end{array} 
 \right) = \left\{ \left( 
\begin{array}{c} 0 \\ 
 \hbar E(x,t) \end{array} 
 \right) + \left[ H_0 + \left( 
 \begin{array}{cc} \hbar g \left(|\phi_{\mathrm{x}} |^2 - \frac{1}{\Delta x}\right) - i\hbar \frac{\gamma_{\mathrm{x}}}{2} & 0 \\ 
   0  & -i\hbar \frac{\gamma_{\mathrm{c}}}{2} \end{array} 
 \right) \right] \left( 
\begin{array}{c} \phi_{\mathrm{x}} \\ 
 \phi_{\mathrm{c}} \end{array} 
 \right)\right\}\,dt +  \frac{\hbar}{\sqrt{4\,\Delta x}}
 \left( 
\begin{array}{c} \sqrt{\gamma_{\mathrm{x}}}\,dW_{\mathrm{x}} \\ 
\sqrt{\gamma_{\mathrm{c}}}\,dW_{\mathrm{c}}\end{array} 
 \right) \, ,
\end{equation}
\end{widetext}
where $\Delta x$ is the spacing of the real-space grid and $dW_{\mathrm{x}}$ and $dW_{\mathrm{c}}$ are complex valued, zero-mean, independent Gaussian noise terms, with white noise correlation in both space and time
\begin{equation}
\overline{dW^*_{i}(x,t)\,dW_{j}(x',t)}=2 \delta_{{x},{x}'}\,\delta_{ij}\,dt
\end{equation}
with $i,j=\{{\rm x},{\rm c}\}$.
In the numerical simulations, we reconstruct the equilibrium Wigner distribution by first letting the system evolve to its steady state under the monochromatic pump, and then taking a large number of independent configurations by sampling the stochastic evolution at different times spaced by $T_{\rm sam}$. In order to ensure statistical independence of the different realizations, a large enough $T_{\rm sam}$ has to be taken such that $T_{\rm sam} \gamma_{{\rm x},{\rm c}} \gg 1$. In practice, a number of realization on the order of $10^5$ is used, taken at time intervals on the order of $T_{\rm sam}= 10 \,\gamma^{-1}_{{\rm x},{\rm c}}$.
As usual in Wigner approaches~\cite{QO,carusott2005prb}, the stochastic averages over the configurations of different functions of the fields provide the expectation value of the corresponding symmetrically ordered operator.

Inspired from previous numerical studies of acoustic black holes in atomic condensates~\cite{carusott2008njp}, we shall look for the signature of analog Hawking radiation in the normalized, zero-delay correlation of the cavity field intensity defined as
\begin{equation}\label{eq:g2def}
 g_{\mathrm{c}}^{(2)}(x,x') = 
 \frac{\langle \hat{\psi}^{\dagger}_{\mathrm{c}} (x) \hat{\psi}^{\dagger}_{\mathrm{c}} (x') 
                        \hat{\psi}_{\mathrm{c}} (x') \hat{\psi}_{\mathrm{c}} (x)  \rangle}
 {\langle \hat{\psi}^{\dagger}_{\mathrm{c}} (x) \hat{\psi}_{\mathrm{c}} (x)  \rangle
   \langle  \hat{\psi}^{\dagger}_{\mathrm{c}} (x')  \hat{\psi}_{\mathrm{c}} (x') \rangle }\, ,
\end{equation}
Within the Wigner formalism, the different contributions to \eq{g2def} have the form~\cite{carusott2005prb}
\begin{equation}\label{eq:wigner1}
\langle  \hat{\psi}^{\dagger}_{\mathrm{c}} (x) \hat{\psi}_{\mathrm{c}} (x)  \rangle =
        \langle |\phi_{\mathrm{c}} (x)|^2 \rangle_W -\frac{1}{2 \,{\Delta}x}      
\end{equation}
and
\begin{multline}
\label{eq:wigner2}
\langle  \hat{\psi}^{\dagger}_{\mathrm{c}} (x) \hat{\psi}^{\dagger}_{\mathrm{c}} (x') 
                        \hat{\psi}_{\mathrm{c}} (x') \hat{\psi}_{\mathrm{c}} (x)  \rangle = \\
             = \left\langle |\phi_{\mathrm{c}} (x)|^2 |\phi_{\mathrm{c}} (x')|^2 \right\rangle_W + \frac{1}{4 \,{\Delta}x^2} (1+\delta_{x,x'})+ \\
             -\frac{1}{2 \,{\Delta}x}(1+\delta_{x,x'})\left\langle |\phi_{\mathrm{c}} (x)|^2 + |\phi_{\mathrm{c}} (x')|^2 \right\rangle_W          \, ,
\end{multline}
where the $\langle \ldots \rangle_W$ averages indicate the classical averages over the different stochastic configurations of the field sampled at
interval times $T_{\rm sam}$. 

\section{Creating the acoustic black hole}\label{horizon}

After having reviewed the technical tools to study the dynamics of the polariton quantum fluid, we can now proceed to discuss realistic configurations that can be used to generate an analog acoustic black hole with a large surface gravity.  All simulations are performed by solving the generalized Gross-Pitaevskii equation \eq{mf_gpe} using experimental parameters taken from the recent work in Ref.~\onlinecite{wertz2010}.

\begin{figure}[t]
 \begin{center}
  \includegraphics[width=0.46\textwidth]{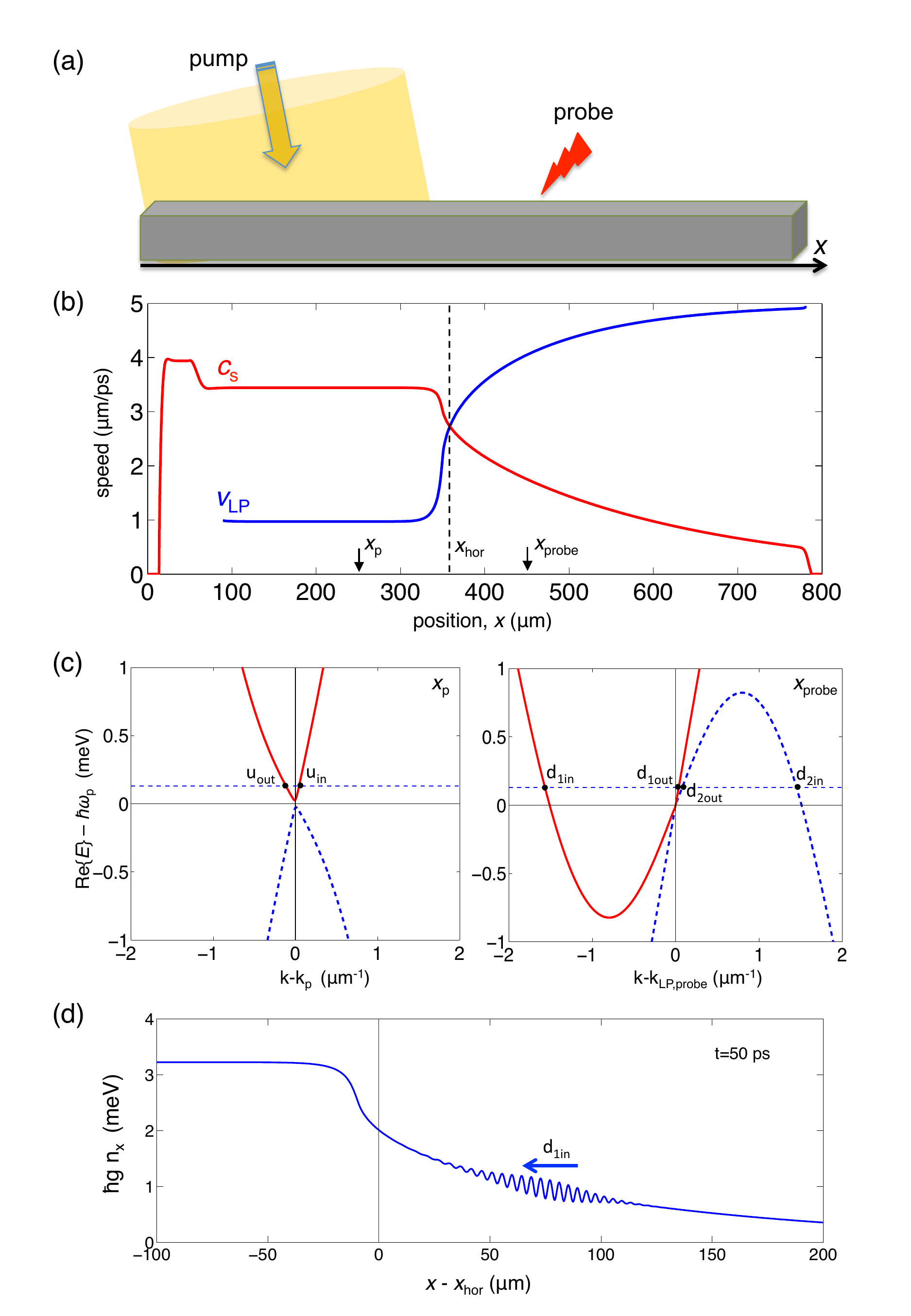}
  \caption{(Color online)   
(a) Scheme of the purely ballistic flow configuration to create a flowing polariton superfluid with an acoustic horizon. (b) Spatial profiles of the flow $v_{\mathrm{LP}}$ and the sound $c_{\mathrm{s}}$ speeds at steady state. The parameters and the spatial shape of the pump laser are the same as in Fig.~\ref{fig2}. 
(c) Dispersion of the collective excitations at two spatial positions: point $x_{\rm p}$ (left panel) is chosen to be in the sub-sonic flat-top region; point $x_{\rm probe}$ (right panel) is located downstream of the horizon in the super-sonic region. Red solid (blue dashed) lines refer to positive (negative) norm Bogoliubov modes.
(d) Spatial profile of the exciton density at a time $t=50$~ps after the arrival of the probe pulse.
 The  probe laser parameters are chosen so to resonantly excite the  Bogoliubov branch indicated as $d_{1,in}$ in panel (c): frequency $\hbar (\omega_{\mathrm{s}} - \omega_{\mathrm{p}}) = 0.13$ meV,  wavevector $k_{\mathrm{s}} - k_{LP}(x_{\mathrm{probe}}) = -1.55\,\mu\textrm{m}^{-1}$. The probe spot has a gaussian shape centered at $x_{\rm probe}=450\,\mu$m with a waist $w_s=20\,\mu$m and a peak amplitude $\hbar F_{\mathrm{s}}=0.35$ meV$\cdot\mu$m$^{-1/2}$. Its temporal shape is also gaussian of duration $\tau = 10$ ps. All numerical calculations are performed in a 800-$\mu$m simulation box with 2048 lattice points and absorbing boundary conditions at the edges of the box. }
     \label{fig:noBH_scheme}
 \end{center}
 \end{figure}

The most straightforward configuration to generate an acoustic horizon was proposed by Solnyshkov et al.~\cite{malpuech2011prb} and is illustrated in Fig.~\ref{fig:noBH_scheme}(b): the presence of losses is responsible for a spatially decreasing density profile of a ballistically flowing polariton condensate and a correspondingly increasing velocity profile. Eventually, the polariton density tends to zero and the flow becomes necessarily super-sonic. The polariton density and speed in the pumped region is fixed by the pump beam parameters: if these are suitably chosen to have an initially sub-sonic flow, a horizon must necessarily appear at some point. 

In spite of the simplicity of this configuration, some care has to be paid to the value of Hawking temperature that can be expected. For smooth horizons within the hydrodynamic limit~\cite{LivRev,carusott2008njp}, the Hawking temperature $T_H$ is determined by the surface gravity 
\begin{equation}
\kappa\equiv\left.\frac{1}{2c_{s}(x)} \frac{d}{dx}[v_{LP}^2(x)-c_{s}^2(x)] 
\right|_{x_{\rm hor}}
 \eqname{surf_grav}
\end{equation}
according to $T_H ={\hbar\kappa}/{k_B}$, with $k_B$ the Boltzmann constant. According to our numerical simulations and analytical approximations~\cite{kamchatnov2011}, the characteristic length scale of the variation of the flow parameters (density and flow speed) is inversely proportional to the loss rate $\ell\propto v_{LP}/\gamma_{LP}$. Direct combination of these two general facts shows that the achievable values of the Hawking temperature is limited from above by the loss rate: for the case in the figure, one indeed has $\hbar\kappa \simeq 0.04$ meV. 

\begin{figure}[t]
 \begin{center}
  \includegraphics[width=0.47\textwidth]{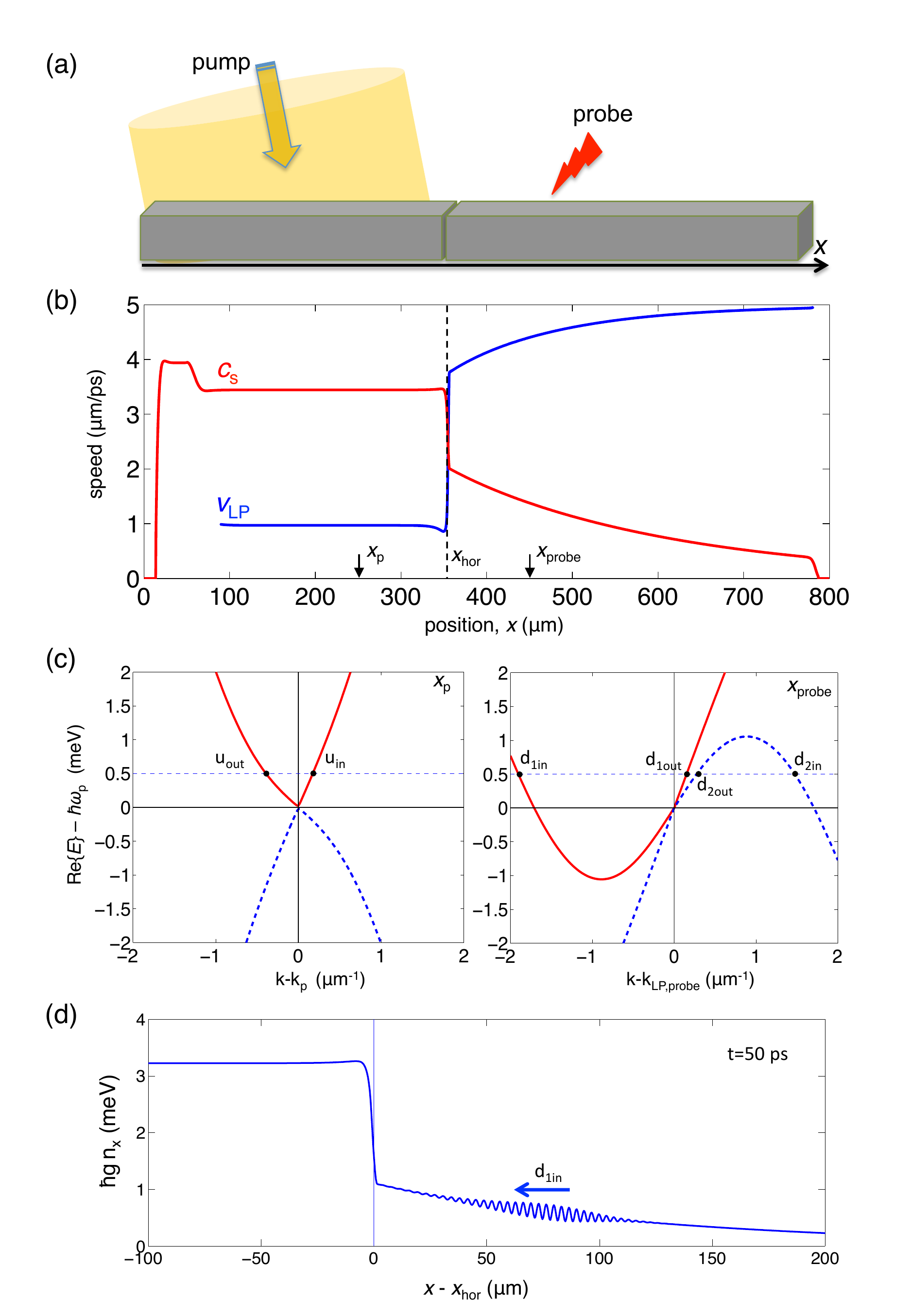}
  \caption{(Color online)   
 (a) Scheme of the ballistic flow configuration with a defect so to create an acoustic horizon with a large surface gravity.
 Panels (b-d) are analogous to the corresponding ones of Fig.~\ref{fig:noBH_scheme}. The parameters and the spatial shape of the pump laser are again the same as in Fig.~\ref{fig2}. The defect consists of a square potential barrier of thickness $L_{\mathrm{def}}=2.5\,\mu$m and is located at a position $x_{\mathrm{def}}=354\,\mu$m close to the flat-top edge. The height of the defect potential is $V_{\mathrm{def}}=1$ meV and acts on both exciton and photon components of the polariton field. The horizon appears in the close vicinity of the defect with a large surface gravity.
The probe laser  parameters are chosen so to resonantly excite the Bogoliubov branch indicated as $d_{1,in}$ in panel (c): frequency $\hbar (\omega_{\mathrm{s}} - \omega_{\mathrm{p}}) = 0.5$ meV, wavevector  $k_{\mathrm{s}} - k_{LP}(x_{\mathrm{probe}}) = -1.9\,\mu\textrm{m}^{-1}$. The shape and the amplitude of the pump spot are the same as in Fig.~\ref{fig:noBH_scheme}(d).       }
       \label{fig:BH_scheme}
 \end{center}
 \end{figure}

A possible solution to overcome this difficulty, is to insert a narrow repulsive potential in the ballistic flow region, close to the edge of the pump spot. The efficiency of this technique is illustrated by the numerical calculations shown in Fig.~\ref{fig:BH_scheme}(b).
Flow across the defect mostly occurs via tunneling processes through the potential barrier: this produces a sudden drop of the condensate density and a corresponding sudden increase of the flow speed. As a result, a horizon appears in the vicinity of the defect center, separating a sub-sonic upstream region from a super-sonic downstream one, with a surface gravity as high as $\hbar\kappa\simeq 1.2$ meV, 
i.e. about a factor $\sim 30$ larger than the purely ballistic flow case of Fig.~\ref{fig:noBH_scheme}.

Another difficulty of the purely ballistic flow configuration of Fig.~\ref{fig:noBH_scheme}(b) was the significant spatial variation of the polariton density on both sides of the horizon, which may strongly distort the geometrical structure Hawking signal.
The defect configuration of Fig.~\ref{fig:BH_scheme}(b) appears to be favorable also in this respect: as the flow in the upstream region is sub-sonic, no Bogoliubov-\u Cerenkov emission can take place from the defect in this direction~\cite{ICCC_RMP} and the polariton density remains almost flat across the whole flat-top region. The slow spatial variation of the density in the downstream super-sonic region is hardly avoided unless more complex laser configurations are used, but does not seem to prevent identification of the Hawking effect. 

On the other hand, the use of a coherent pump raises the crucial issue of avoiding all those branch-sticking and gap effects typical of non-equilibrium condensates~\cite{carusotto2004prl,ciuti2005pss} that would spoil the analogy with gravitational systems. As we have reviewed in Sec.~\ref{sec:bogoliubov}, choosing the value of the laser amplitude in the flat-top region in the vicinity of the end-point of the upper branch of the hysteresis loop is enough to have a linear, sonic dispersion of the low-$k$ dispersion of the form \eq{sonic} at all spatial positions around the horizon~\footnote{In practical simulations, we keep the pump amplitude at a value slightly above the turning point to allow a faster and more robust convergence towards the steady state. Even if the resulting Bogoliubov dispersion shows a small gap, its amplitude is however much smaller than the polariton broadening $\gamma_{LP}$, and hence has no influence on the final results, which further supports the robustness of our scheme.}. 

As it was illustrated in Fig.~\ref{fig2}, keeping the system stably on the upper branch of the hysteresis loop requires including a region of higher pump intensity upstream of the flat-top. Exception made for its role in switching on the optical bistability, the high pump intensity region play no role in the Hawking physics. As we have discussed in Sec.~\ref{sec:bogoliubov}, the Bogoliubov dispersion in the downstream region is guaranteed to be sonic at low wavevectors by the ballistic nature of the flow in this region. The flow speed is guaranteed to be super-sonic by the suppressed value of the density beyond the defect.

\begin{figure}[b]
 \begin{center}
 \vspace{0cm}
  \includegraphics[width=0.5\textwidth,clip]{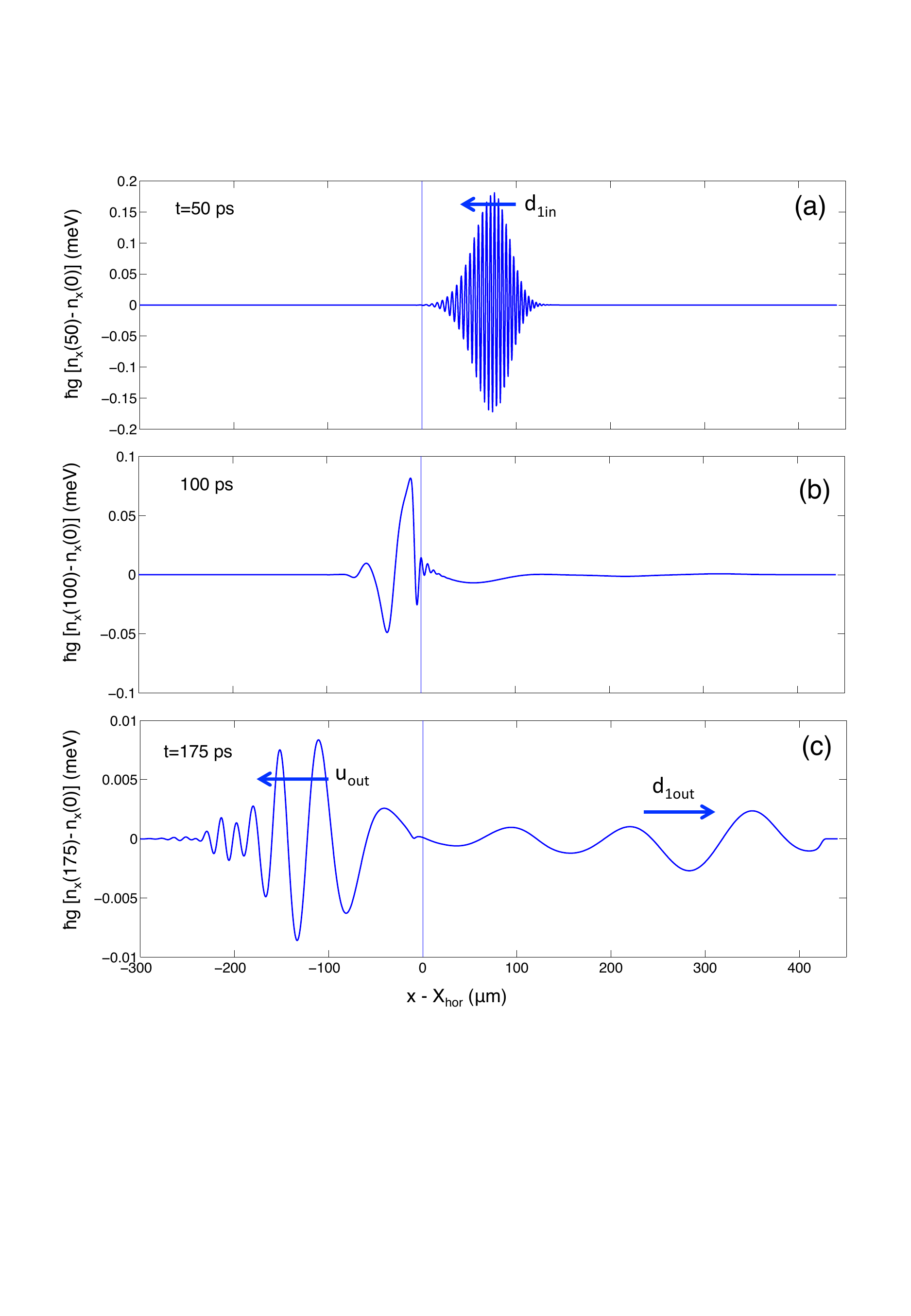}
  \vspace{-2cm} 
  \caption{(Color online) Snapshots of the spatial profile of the density modulation due to the propagating phonon wavepackets at different times (a) $t=50$ ps, (b) $t=100$ ps, and (c) $t=175$ ps after the arrival of the probe pulse. Each panel displays the excitonic density modulation with respect to the time-independent steady state.  
  The pump and probe parameters are the same as for the purely ballistic flow shown in Fig.~\ref{fig:noBH_scheme}(d). The labels on the wavepackets refer to the Bogoliubov dispersions of Fig.~\ref{fig:noBH_scheme}(c), while the arrows indicate the propagation direction of each wave packet. }
      \label{fig:noBH_snap}
 \end{center}
 \end{figure}

\begin{figure}[t]
 \begin{center}
  \includegraphics[width=0.5\textwidth,clip]{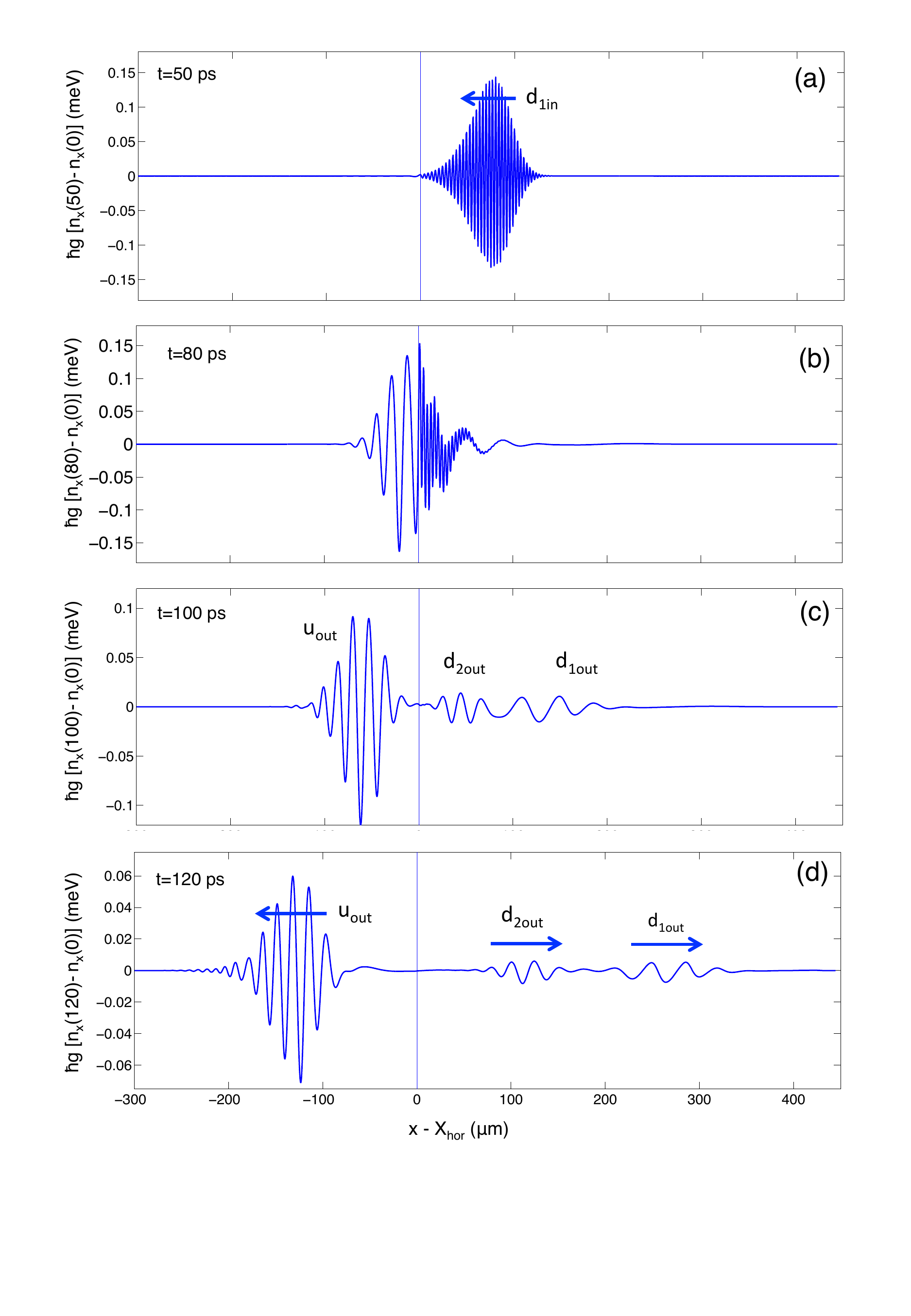}
  \vspace{-2.0cm} 
  \caption{(Color online)     Snapshots of the spatial profile of the density modulation due to the propagating phonon wavepackets at different times (a) $t=50$ ps, (b) $t=100$ ps, and (c) $t=175$ ps after the arrival of the probe pulse.  Each panel displays the excitonic density modulation with respect to the time-independent steady state. The pump and probe parameters correspond to the defect configuration shown in Fig.~\ref{fig:BH_scheme}(d). The labels on the wavepackets refer to the Bogoliubov dispersions of Fig.~\ref{fig:BH_scheme}(c), while the arrows indicate the propagation direction of each wave packet. }
     \label{fig:BH_snap}
 \end{center}
 \end{figure}

\section{Wavepacket scattering}
\label{scattering}

The numerical simulations that we have presented in the previous section demonstrate the possibility to create configurations of a polariton flow through a horizon separating regions of sub- and super-sonic regimes with a sizable surface gravity. This configuration will be a powerful workbench for studies of Hawking phenomena. Inspired by related theoretical work on atomic condensates~\cite{recati2009pra} and experiments in classical hydrodynamics~\cite{unruh_water}, in this Section we shall first consider the scattering of a phonon wavepacket off the horizon, a process where the classical counterpart of the Hawking effect manifests itself as an additional reflected wavepacket. This discussion will be the starting point for the next section, where we shall investigate the very Hawking effect, namely the conversion of zero-point quantum fluctuations into observable radiation by the horizon.

The scattering dynamics of Bogoliubov phonons hitting the horizon can be studied in terms of the generalized GPE \eq{mf_gpe}: far from the horizon, the propagation of weak wavepacket perturbations follows the Bogoliubov dispersion discussed in Sec.~\ref{sec:bogoliubov} albeit with spatial dependent flow and sound speeds. Figs.~\ref{fig:noBH_scheme} to \ref{fig:BH_snap} illustrate the different aspects of this physics for the two previously mentioned cases, namely a smooth flow with a low surface gravity a defect configuration with a sizable surface gravity. For each configuration, the structure of the flow is illustrated in Figs.~\ref{fig:noBH_scheme}(b) and \ref{fig:BH_scheme}(b), while the smaller (c,d) panels show the Bogoliubov dispersion at different spatial positions.

Once the system is in its steady state, a low-frequency wavepacket perturbation is generated in the downstream super-sonic region by means of an extra coherent laser pulse with a spatio-temporal shape of the form
\begin{equation}
E(x,t)=F_s\,e^{-(x-x_{s})^2/w_s^2}\,e^{-(t-t_s)^2/\tau^2}\,e^{i(k_s x - \omega_s t)}\, .
\end{equation}
The probe wave vector ($k_s$) and frequency ($\omega_s$) are chosen to be resonant with the Bogoliubov branch labeled as $d_{1,in}$ in the (c) panels. The probe beam is centered at $x_{\rm probe}$; its waist $w_s$ determines the spatial size of the generated wavepacket (and its inverse width in $k$-space). The temporal duration $\tau$ has to be long enough not to excite the upper polariton branch nor the negative-norm Bogoliubov branch. Its amplitude $F_s$ is chosen weak enough to remain within the validity domain of a linearized description of excitations.

The generated wavepacket shown in Figs.~\ref{fig:noBH_scheme}(d) and \ref{fig:BH_scheme}(d) then propagates in the leftward direction against the horizon. A series of snapshots of this evolution are presented in Figs.~\ref{fig:noBH_snap} and \ref{fig:BH_snap}.
The difference between the behaviors of the weak and of a large surface gravity cases is apparent: in the former case, only two wavepackets visibly emerge from the horizon at long times, located on the $u_{out}$ and $d_{1,out}$ Bogoliubov branches. On the other hand, in the latter case, an extra wavepacket appears on the $d_{2,out}$ branch. Remarkably, the Bogoliubov norm of this wavepacket is negative, which signals the occurrence of Hawking conversion processes. The identification of the different wavepackets in terms of their Bogoliubov branch is confirmed by the numerical measurement of their group velocity and of their wavevector. 

Of course, the physical reason why we did not observe a $d_{2,out}$ wavepacket in Fig.~\ref{fig:noBH_snap} is that the low value of the surface gravity strongly suppresses the amplitude of the Hawking process. As it was expected on general grounds~\cite{LivRev} and then explicitly verified~\cite{parentani,scott2012}, the Hawking scattering amplitude for a Bogoliubov wavepacket of given carrier frequency $\omega$ scales down exponentially with the surface gravity according to the Boltzmann factor $\exp(-2\pi \omega/\kappa)$.
On the other hand, excitation of the $d_{2,out}$ branch would be totally forbidden if the flow was everywhere sub-sonic and no horizon was present: in this case, in fact, energy could not be conserved in the Hawking scattering process.

\section{Spatial correlations of density fluctuations}
\label{Hawking}

\begin{figure}[t]
 \begin{center}
  \includegraphics[width=0.5\textwidth]{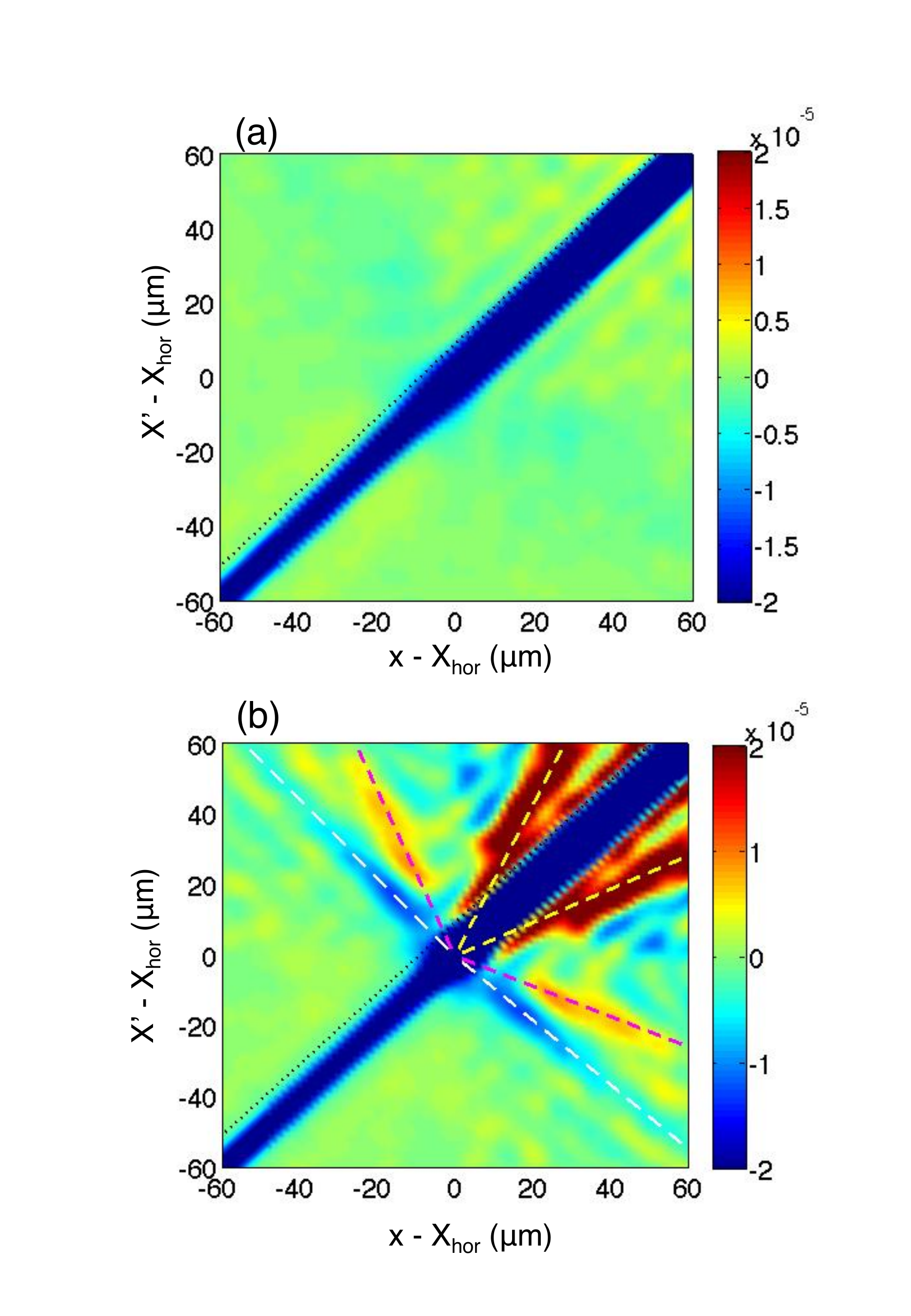}
  \vspace{-0.5cm} 
  \caption{(Color online)    Color scale plot of the normalized spatial correlation function of photon density fluctuations $g_c^{(2)}(x,x')$. This quantity directly reflects in the intensity correlations of a near-field image of the emitted light from the microcavity.
  The upper (a) panel refers to the case of a purely ballistic flow considered in Figs.~\ref{fig:noBH_scheme} and \ref{fig:noBH_snap}. The lower panel (b) refers to the case with a defect in the ballistic flow considered in Figs.~\ref{fig:BH_scheme} and \ref{fig:BH_snap}. The colored dashed lines in (b) indicate the different Hawking features as discussed in the text.}
     \label{fig:g2}
 \end{center}
 \end{figure}

The process studied in the previous section provides a solid evidence of the classical counterpart of the Hawking effect, namely the possibility of inter-converting positive and negative norm waves at the horizon. Now we shall proceed with the study of the Hawking effect {\em stricto sensu}, i.e. the emission of radiation by the horizon via the inter-conversion of zero-point quantum fluctuations into observable radiation.

To this purpose, one has to fully include in the model the quantum fluctuations of the fields around the mean-field value predicted by the Gross-Pitaevskii equation. In this work, we adopt the truncated Wigner approach described in Sec.~\ref{sec:montecarlo}, where expectation values of quantum operators are calculated as classical averages over a suitably chosen stochastic partial differential equation.
Inspired from previous theoretical~\cite{balbinot2008pra} and numerical~\cite{carusott2008njp} work, we shall consider the spatial correlation function of the quantum fluctuations of the fluid density: the smoking gun of Hawking radiation consist of a negative long-range correlation signal between points located on opposite sides of the horizon.

Numerical plots of the normalized density-density correlation function $g^{(2)}_c(x,x')-1$ are shown in the two panels of Fig.~\ref{fig:g2} for the two cases of a weak and strong surface gravity $\kappa$, respectively.
While in the upper panel, Fig.~\ref{fig:g2}(a) (weak surface gravity, $\hbar\kappa\approx 0.04$ meV) there is no clear feature emerging over noise, the lower panel, Fig.~\ref{fig:g2}(b) (strong surface gravity, $\hbar\kappa\approx 1.2$ meV) shows several features that can be interpreted as a direct consequence of the analog Hawking radiation~\cite{recati2009pra,parentani}: the two Bogoliubov excitations forming the Hawking pair are simultaneously emitted from the horizon. At later times, the temporal correlation between the emission time of the Hawking partners reflects into a long-distance correlation in the density fluctuations due to each member of the pair.
While in the atomic case this correlation extends  at long times to indefinitely large distances from the horizon, the finite decay rate of Bogoliubov excitations in polariton fluids predicted in \eq{sonic} restricts the correlation signal to a finite region of size $v_{\rm g}/\gamma_{LP}$ around the horizon, $v_{\rm g}$ being the group velocity of the Bogoliubov branch under consideration. Away from this region around the horizon, the Hawking phonons have decayed and the corresponding density fluctuations have disappeared.
The dashed lines in Fig.~\ref{fig:g2}(b) indicate the location at which one would expect the maximal density correlations due to the quantum vacuum emission process into the $u_{out}-d_{2,out}$ (white), the $u_{out}-d_{1,out}$ (purple), the $d_{1,out}-d_{2,out}$ (yellow) pair of modes.

The first process is the traditional Hawking emission process where the two quanta are emitted into opposite directions from the horizon. The main properties of the density correlation pattern due to this process is well captured by the non-dispersive quantum field theory on curved space-time as proposed in Ref.~\onlinecite{balbinot2008pra}. For $x-x_{\rm hor}>0$ and $x'-x_{\rm hor}<0$, the correlation is peaked along the half-line
\begin{equation}
\frac{x-X_{\rm hor}}{v_d - c_d}=\frac{x'-X_{\rm hor}}{v_u-c_u}.
\eqname{ud2}
\end{equation}
Here, $v_{d,u}$ and $c_{d,u}$ are the flow speed and speed of sound in the upstream and downstream regions, so that $v_d-c_d$ and $v_u-c_u$ are the group velocity of the $d_{2,out}$ and $u_{out}$ modes in the low-$k$, sonic region. While in the atomic BEC case of Ref.~\onlinecite{carusott2008njp} their value did not depend on space in the asymptotic regions away from the horizon, the driven-dissipative nature of polaritons is responsible for the spatial dependence of the flow $v_{LP}$ and sound $c_s$ speeds that is visible in Fig.~\ref{fig:BH_scheme}(b), mostly in the downstream region. However, as the space dependence remains moderate within the regions of thickness $|v_u-c_u|/\gamma_{LP}$ and $|c_d-v_d|/\gamma_{LP}$ on either side of the horizon where correlations are significant, no appreciable curvature of the Hawking tongues is visible in Fig.~\ref{fig:g2}(b).

The orientation of the purple and yellow dashed lines in Fig.~\ref{fig:g2}(b) is analogously obtained from the group velocity of the corresponding modes, namely
\begin{equation}
\frac{x-X_{\rm hor}}{v_d + c_d}=\frac{x'-X_{\rm hor}}{v_u-c_u}
\eqname{ud1}
\end{equation}
and
\begin{equation}
\frac{x-X_{\rm hor}}{v_d - c_d}=\frac{x'-X_{\rm hor}}{v_d+c_d}
\eqname{d1d2}
\end{equation}

The excellent agreement of the geometrical location of the numerically observed features with the analytical predictions (\ref{eq:ud2}--\ref{eq:d1d2}) confirms our interpretation.
Remarkably~\footnote{
While the qualitative structure of the Hawking correlation pattern appears quite similar, quantitative comparison of our results with the ones in Ref.~\onlinecite{malpuech2011prb} is hardly made, as the scale of the correlation signal is missing there. Differently from that paper, the negative sign of the correlation signal reported in the present work matches the one expected from theoretical and numerical work in equilibrium condensates~\cite{balbinot2008pra,carusott2008njp}. This sign difference is most likely due to the fact that in Ref.~\onlinecite{malpuech2011prb} fluctuations are induced by some classical disorder present in the super-sonic region.
},
the peak value of the $u_{out}-d_{2,out}$ Hawking correlation signal is on the order of $g^{(2)}_{{\rm c}}-1\simeq 0.92 \times 10^{-5}$, which is not too far from the analytical prediction of Ref.~\onlinecite{balbinot2008pra}. As the surface gravity is here comparable to the interaction energy $\hbar g_{LP} n_{LP}$, the quantitative discrepancy can be easily traced back to the breakdown of the hydrodynamical approximation underlying the analytical model~\cite{carusott2008njp}. Thanks to the continuous-wave nature of the proposed experimental setting, the quantitatively low value signal to be observed can be overcome by a sufficiently long integration time.

While agreement with the atomic case is good also for the $u_{out}-d_{1,out}$ feature, the $d_{1,out}-d_{2,out}$ one has a different sign. A possible explanation of this behavior can be traced back to the non-universality of the $u_{out}\rightarrow d_{1,out}$ back-scattering effect that is responsible for the conversion at the horizon of the standard $u_{out}-d_{2,out}$ Hawking correlation into the $d_{1,out}-d_{2,out}$ one.

Before concluding, it is important to note that throughout the whole discussion, we have implicitly assumed that the $g_c^{(2)}(x,x')$ correlation function is assumed to be measured at the same time: a rough estimate of the required temporal resolution of the detector is given by $\delta t\approx \delta x/c_{u,d}$, where $c_{u,d}$ is the speed of sound in the polariton gas (on the order of $2\,\mu$m/ps from Figs.~\ref{fig:BH_scheme}) and $\delta x$ is the spatial width of the correlation signal. Using the value $\delta x\approx 8\,\mu$m taken from Fig.~\ref{fig:g2}, one can estimate the needed temporal resolution to be on the order of $\delta t\approx 4\,$ps, which is within the state-of-the-art of optical technology.

\section{Conclusions}\label{concl}

In this article we have presented a comprehensive study of classical and quantum hydrodynamic properties of acoustic black holes in superfluids of exciton-polaritons in semiconductor microcavities. Inspired from on-going experimental research, we have identified and characterized one-dimensional polariton wire devices as model systems, where the polaritons injected by a single monochromatic laser beam with a suitable spot profile generate a flow configuration showing an acoustic black hole horizon. Inserting a repulsive potential defect provides a dramatic enhancement of the analog surface gravity at the horizon.  Even if our calculations have been performed for the specific case of a semiconductor microcavity in the strong-coupling regime, all conclusions straightforwardly extend to a generic planar cavity filled by a nonlinear optical medium, as originally suggested in Ref.~\onlinecite{marino2008}, or photonic crystal polariton states with engineered mode dispersion~\cite{gerace2007,bajoni2009}.

In analogy with previous works on acoustic black holes in atomic condensates and classical hydrodynamics of surface waves, two experiments have been proposed and numerically simulated to assess the efficiency of Hawking wave-conversion processes at the horizon.
When a coherent wavepacket of Bogoliubov phonons is incident on the horizon, the Hawking effect is visible as the production of an extra emerging wavepacket on the negative-norm branch. This is the classical evidence for the Hawking process taking place at the horizon.
On the other hand, the usual Hawking radiation is a purely quantum effect consisting of the conversion of zero-point fluctuations into a stream of observable phonons at the horizon: the correlation between the emission times of a phonon and its Hawking partner is visible as a long-distance correlation between the density fluctuations on either side of the horizon. The density correlations provide the clearest experimental signature of Hawking radiation in superfluids: in the present polariton case, in-cavity density correlations directly transfer into the secondary emission from the cavity photon, and can be observed as analogous correlations in the intensity fluctuations of the near-field emission. As soon as a sufficient signal over noise ratio is achieved, an experiment along the proposed lines will hopefully provide a clear evidence of a fundamental effect of quantum field theory that so far has eluded all experimental observation, overcoming the barrier to direct observation imposed by the negligibly small Hawking temperature of astrophysical black-holes.

\begin{acknowledgments}
We are grateful to A. Amo, R. Balbinot, J. Bloch, A. Bramati, S. Finazzi, S. Robertson, D. Sanvitto for continuous stimulating discussions, and D. Sarchi for a fruitful collaboration at an early stage of the work. 
DG warmly acknowledges the friendly hospitality of the INO-CNR BEC Center in Trento, where most of this work has been conceived and carried out, and L.C. Andreani for supporting this research.
\end{acknowledgments}

\end{document}